\documentclass[dvipsnames,format=sigconf,anonymous=false,review=false]{acmart}

\AtBeginDocument{%
  \providecommand\BibTeX{{%
    \normalfont B\kern-0.5em{\scshape i\kern-0.25em b}\kern-0.8em\TeX}}}

\copyrightyear{2024}
\acmYear{2024}
\setcopyright{acmlicensed}\acmConference[GECCO '24 Companion]{Genetic
and Evolutionary Computation Conference}{July 14--18, 2024}{Melbourne, VIC,
Australia}
\acmBooktitle{Genetic and Evolutionary Computation Conference (GECCO '24
Companion), July 14--18, 2024, Melbourne, VIC, Australia}
\acmDOI{10.1145/3638530.3664107}
\acmISBN{979-8-4007-0495-6/24/07}

\usepackage{blindtext}
\usepackage{braket}
\usepackage{nicefrac}
\usepackage{mathtools}

\DeclareMathOperator*{\argmin}{arg\,min}
\DeclareMathOperator*{\diag}{diag}

\usepackage{amsmath}
\usepackage{outlines}
\usepackage{mathtools}
\usepackage{graphicx}
\usepackage{caption,mwe}
\usepackage{subcaption}

\begin{document}

\title{Solving the Turbine Balancing Problem using Quantum Annealing}

\author{Arnold Unterauer}
\orcid{0009-0009-6758-1093}
\affiliation{%
  \institution{Aqarios GmbH}
  \streetaddress{Prinzregentenstraße 120}
  \city{Munich}
  \state{Bavaria}
  \country{Germany}
  \postcode{81677}}
  \email{arnold.unterauer@aqarios.com}

\author{David Bucher}
\orcid{0009-0002-0764-9606}
\affiliation{%
  \institution{Aqarios GmbH}
  \streetaddress{Prinzregentenstraße 120}
  \city{Munich}
  \state{Bavaria}
  \country{Germany}
  \postcode{81677}}
\email{david.bucher@aqarios.com}

\author{Matthias Knoll}
\orcid{0009-0007-2767-7969}
\affiliation{%
  \institution{Aqarios GmbH}
  \streetaddress{Prinzregentenstraße 120}
  \city{Munich}
  \state{Bavaria}
  \country{Germany}
  \postcode{81677}}
\email{matthias.knoll@aqarios.com}

\author{Constantin Economides}
\orcid{0009-0004-1022-7249}
\affiliation{%
  \institution{Aqarios GmbH}
  \streetaddress{Prinzregentenstraße 120}
  \city{Munich}
  \state{Bavaria}
  \country{Germany}
  \postcode{81677}}
\email{constantin.economides@aqarios.com}

\author{Michael Lachner}
\orcid{0009-0008-6874-8329}
\affiliation{%
  \institution{Aqarios GmbH}
  \streetaddress{Prinzregentenstraße 120}
  \city{Munich}
  \state{Bavaria}
  \country{Germany}
  \postcode{81677}}
\email{michael.lachner@aqarios.com}

\author{Thomas Germain}
\orcid{0009-0007-0289-1686}
\affiliation{%
  \institution{MTU Aero Engines AG}
  \streetaddress{Dachauer Straße 665}
  \city{Munich}
  \state{Bavaria}
  \country{Germany}
  \postcode{80995}}
\email{thomas.germain@mtu.de}

\author{Moritz Kessel}
\orcid{0009-0002-3423-6532}
\affiliation{%
  \institution{MTU Aero Engines AG}
  \streetaddress{Dachauer Straße 665}
  \city{Munich}
  \state{Bavaria}
  \country{Germany}
  \postcode{80995}}
\email{moritz.kessel@mtu.de}

\author{Smajo Hajdinovic}
\orcid{0009-0004-0381-1957}
\affiliation{%
  \institution{MTU Aero Engines AG}
  \streetaddress{Dachauer Straße 665}
  \city{Munich}
  \state{Bavaria}
  \country{Germany}
  \postcode{80995}}
\email{smajo.hajdinovic@mtu.de}

\author{Jonas Stein}
\orcid{0000-0001-5727-9151}
\affiliation{%
  \institution{Aqarios GmbH}
  \streetaddress{Prinzregentenstraße 120}
  \city{Munich}
  \state{Bavaria}
  \country{Germany}
  \postcode{81677}}
\email{jonas.stein@aqarios.com}

\renewcommand{\shortauthors}{Unterauer et al.}

\begin{abstract}
Quantum computing has the potential for disruptive change in many sectors of industry, especially in materials science and optimization. In this paper, we describe how the Turbine Balancing Problem can be solved with quantum computing, which is the NP-hard optimization problem of analytically balancing rotor blades in a single plane as found in turbine assembly. Small yet relevant instances occur in industry, which makes the problem interesting for early quantum computing benchmarks. We model it as a Quadratic Unconstrained Binary Optimization problem and compare the performance of a classical rule-based heuristic and D-Wave Systems' Quantum Annealer Advantage\_system4.1. In this case study, we use real-world as well as synthetic datasets and observe that the quantum hardware significantly improves an actively used heuristic's solution for small-scale problem instances with bare disk imbalance in terms of solution quality. Motivated by this performance gain, we subsequently design a quantum-inspired classical heuristic based on simulated annealing that achieves extremely good results on all given problem instances, essentially solving the optimization problem sufficiently well for all considered datasets, according to industrial requirements.

\end{abstract}





\maketitle

\section{Introduction}
\label{sec:introduction}
The Turbine Balancing Problem (TBP) is an NP-hard combinatorial optimization problem \cite{Amiouny1993} central to the assembly process of turbines. It describes the problem of assigning the positions of a set of blades to fixed positions (called \emph{slots} in the remainder of this paper) of the rotor disk or hub to minimize the imbalance resulting from differently weighted blades caused by imperfect fabrication. \emph{Static imbalance}, i.e., any deviation from the turbine's center of gravity to the axis of rotation, causes vibrations and stress on the bearings and thus shortens the effective lifespan of the turbine. In practice, manufacturers typically compensate this imbalance by adding weights to the turbine fan, which is also called \emph{static balancing}~\cite{Anon90}. As any additional weight is undesirable in aircraft engines, and to reduce time spent in the balancing task, many heuristic approaches to the TBP have been developed~\cite{Amiouny2000, MASON1997153, CHOI20041245, sym13050832}.

A less investigated case is the bare imbalance of the rotor disk, which can significantly contribute to the resulting imbalance. Many actively utilized heuristics must be explicitly extended to account for it. In practice, these considerations demand for time consuming, manual adjustments which in turn slow down the manufacturing process and hence increase costs. Moreover, nearly all employed TBP solvers are focused on fast runtime, as this problem typically needs to be solved with real-time performance to not slow down the production process. 
However, as the number of blades on a turbine increase, existing heuristic approaches often struggle to provide sufficiently good solutions (i.e., assignments that do not necessitate subsequent static balancing)~\cite{Amiouny2000}, especially when including the imbalance of the bare rotor disk.

An emerging technology that has shown the potential to solve optimization problems advantageously compared to known heuristics is quantum computing~\cite{PhysRevX.6.031015, doi:10.1126/science.abo6587, PhysRevX.8.031016}. Exploiting the richer, efficient algorithmic toolset provided by the laws of quantum mechanics, existing quantum computers like \emph{Quantum Annealers} already allow to solve small real world optimization problem instances. A key benefit of using a quantum optimization approach like \emph{Quantum Annealing} (QA) (which is the algorithmic procedure a Quantum Annealer executes) is its flexibility in regards to runtime: The longer it can run, the better the results will generally get. The combination of this feature with the prospect of potentially better solution quality motivates us to investigate the performance of quantum optimization techniques for solving the TBP.

Our contributions to this performance evaluation are summarized in the following:
\begin{itemize}
    \item A Quadratic Unconstrained Binary Optimization (QUBO) formulation of the TBP while considering a possible imbalance of the bare disk, being the necessary input for Quantum Annealers.
    \item A case study evaluating the performance of an industrially employed, classical rule-based heuristic (described in Sec. ~\ref{heuristic_rule}), Qbsolv, and the D-Wave Advantage\_system4.1 Quantum Annealer.
    \item The development of a quantum-inspired classical heuristic based on simulated annealing.
\end{itemize}

The following sections are structured into five parts. Sec.~\ref{sec:background} shows how QUBO problems can be solved on Quantum Annealers. Sec.~\ref{sec:maths} describes the process of formulating the TBP in QUBO form. Sec.~\ref{sec:custom_sa} introduces our novel quantum-inspired solver. Sec.~\ref{sec:experiments} displays the conducted experiments and interprets the results. Sec.~\ref{sec:conclusion} ultimately concludes our findings.

\section{Background}
\label{sec:background}
In this section, we introduce the mechanics underlying QA and how formulating an optimization problem as Quadratic Unconstrained Binary Optimization (QUBO) helps with solving it on a quantum computer.

The \emph{adiabatic theorem} in quantum mechanics states that whenever one Hamiltonian transitions to another Hamiltonian, the physical system remains in the instantaneous ground state, as long as it was prepared in the ground state of the initial Hamiltonian and the evolution happens sufficiently slowly~\cite{born1928}. This is essentially the basis of Adiabatic Quantum Computation (AQC), the underlying computation paradigm of Quantum Annealing (QA), which differs fundamentally from the standard gate-based quantum computing model, by conducting time evolution continuously rather than discrete. AQC starts with a Hamiltonian whose ground state can typically be prepared easily, e.g., the $H_I = -\sum_i X_i$ with the ground state $\ket{+}$. Here $X_i$ and in the following $Z_i$ denote the Pauli $\sigma_x$ and $\sigma_z$ operators acting on the qubit with index $i$. In AQC, this Hamiltonian now undergoes a slow change towards an Ising spin-glass Hamiltonian~\cite{farhi2001}, which is described in Eq.~\ref{eq:IsingSpinnGlass}.
\begin{align}\label{eq:IsingSpinnGlass}
    H_P = -\sum_i h_i Z_i - \sum_{i,j} J_{ij} Z_i Z_j,
\end{align}
so that the Hamiltonian describing the time evolution has the form $H(t) = (1-t) H_I + t H_P$, where $t \in [0, 1]$. Given that the evolution is slow enough such that the adiabatic theorem holds, we are guaranteed to arrive in the ground state of $H_P$. As $H_P$ is strictly diagonal, the ground state will not be entangled and can be expressed solely through the $z$-basis. Thus, $H_P$ is purely classical and can also be described through the classical spin values $Z_i \rightarrow s_i \in \{-1, +1\}$.

A convenient property of the Ising spin-glass problem is, that finding its ground state is generally NP-hard~\cite{barahona1982}. Therefore, mapping various hard problems to the Ising model and running QA emerges as a viable optimization technique. Conveniently, the classical Ising spin-glass is isomorphic to QUBO via the mapping $s_i = 2 x_i - 1$, where $x_i$ are the binary variables of the QUBO, i.e., $\argmin_{x\in\left\lbrace 0,1\right\rbrace^n} x^\top Qx$ where $Q\in\mathbb{R}^{n \times n}$. Due to its often more convenient $0/1$ encoding, QUBO has developed to a standard for specifying optimization problems for QC and many optimization problems have already been formulated in that way~\cite{Glover2022,10.3389/fphy.2014.00005}.

Since today's hardware is typically error-prone and one has to perform fast time evolution in order not to lose information to state decoherence, the final state will not fully be in the ground state and the optimal solution is merely obtained probabilistically~\cite{bapst2013}. Therefore, instead of running QA once, we repeatedly sample from the final state and pick the best outcome.

QA also inspired the Quantum Approximate Optimization Algorithm (QAOA)~\cite{farhi2014}, which essentially is a timestep-parameterized Trotterization of the annealing process. This discretization is necessary to conduct the AQC-based quantum optimization in the quantum gate-model. Most notably, quantum gate-based computation models are already capable of solving higher-order problems beyond the quadratic realm (so called \emph{Polynomial} Unconstrained Binary Optimization (PUBO)), as the therefore necessary multi-qubit interactions can be decomposed into single- and two-qubit gates. This has shown promising performance, even for naturally quadratic problems like the TSP, as specific space-efficient binary encodings can increase the order of the objective function as opposed to the more widespread one-hot encodings~\cite{Salehi2022,10.1145/3583133.3596358}.



\section{Problem Formulation}
\label{sec:maths}
%
The problem of placing $N$ blades in $N$ slots, such that the overall imbalance of the turbine assembly is minimized, is an assignment problem. We want to find an optimal assignment $\sigma: \{1,\dotsc,N\} \rightarrow \{1,\dotsc,N\}$, such that the distance $d$, of the center of mass and center of rotation, is minimal. Of course, we can only place a single blade in a certain slot, i.e., the assignment has a permutation character, therefore, $\sigma \in S_N$, where $S_N$ denotes the symmetric group of degree $N$.

In our problem formulation, we assume that the blades are placed perfectly on the same plane of rotation, i.e., we restrict ourselves to a two-dimensional version of the problem, as the three dimensional version would significantly complicate the cost function and hence exceed the pursued proof of concept. Furthermore, similar to~\cite{doi:10.1126/science.220.4598.671}, we do not contemplate the inertial characteristics of the single blades but design the fans as point masses $m_i$ located at a constant radius $r > 0$. Without loss of generality, we set $r = 1$.

The slots are spread around the central disk in equidistant angular steps, i.e., $\Delta \phi = \frac{2\pi}{N}$. Without loss of generality, we can assume that the first slot is positioned at angle $\phi_1 = 0$. The subsequent positions can then be computed as follows
\begin{align}
    \phi_j = (j - 1) \Delta \phi = \frac{2 \pi (j-1)}{N} \quad  \forall j \in \{1,\dotsc,N\}.
\end{align}

The central disk on which the blades are mounted will also have an bare disk imbalance due to inaccuracies in manufacturing. The imbalance can be described by its $x-$ and $y-$components through $\mathbf{y} \in \mathbb{R}^2$. Alternatively, it may also be referred to in polar coordinates as angle $\phi_0$ with mass $m_0$.

\begin{figure}
\centering
\includegraphics[width=0.7\linewidth]{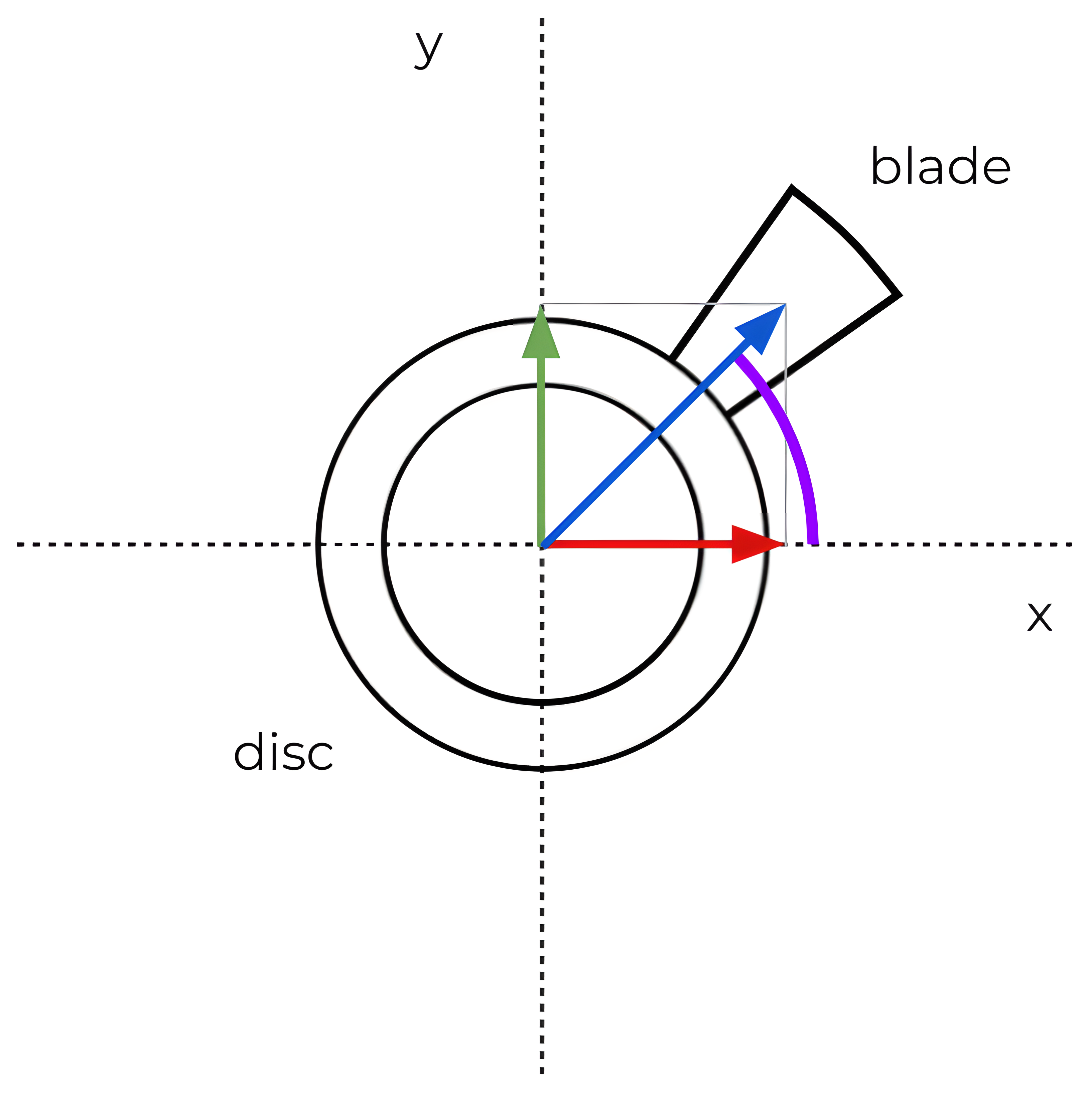}
\caption{Visualization of placing a blade at a certain angle on the disk. The added imbalance (blue arrow) can be split into $x$- and $y$-components (red and green arrows, respectively). }
\label{fig:cart_balance}
\end{figure}

Defining that the center of rotation is at the origin $(0, 0)^\top$, the distance of the center of mass from the origin without any blades attached is $\|\mathbf{y}\|_2$, where $\|\cdot\|_2$ is the Euclidian norm. When mounting blade $i$ with mass $m_i$ at slot $j$, i.e., angle $\phi_j$, the distance becomes
\begin{align}
    d = \left\|\mathbf{y} + m_i \begin{pmatrix}\cos \phi_j \\ \sin \phi_j \end{pmatrix} \ \right\|_2 = \| \mathbf{y} + m_i \mathbf{z}_j\|_2.
\end{align}
A visualization of the above quantity can be seen in Fig.~\ref{fig:cart_balance}.
Considering all blades placed according to the assignment $\sigma$, the objective quantity can be expressed as follows
\begin{align}
    d &= \left\| \mathbf{y} + \sum_{i = 1}^N m_i \mathbf{z}_{\sigma(i)}\right\|_2,
\end{align}
and since $\mathbf{z}_i^\top \mathbf{z}_j = \cos (\phi_i - \phi_j)$ by trigonometric identities, we can simplify the above expression:
\begin{equation}\label{eq:initial_opt}
\begin{split}
    d^2 = m_0^2 &+ 2 m_0 \sum_{i=1}^N m_i \cos(\phi_0 - \phi_{\sigma(i)})  \\ &+ \sum_{i,j = 1}^N m_i m_j \cos(\phi_{\sigma(i)} - \phi_{\sigma(j)}).
\end{split}
\end{equation}

Thus the overall optimization of finding the best placement of the blades is described by the following problem
\begin{align}
    \sigma^* = \argmin_{\sigma \in S_N} d^2(\sigma),
\end{align}
since squaring is monotonic for positive numbers and per definition $d > 0$, choosing the linear or squared distance does not make a difference. However, the squared distance is easier to translate into QUBO formalism, as we will see in the next section.

\begin{figure*}
    \centering
    \includegraphics[width=0.9\textwidth]{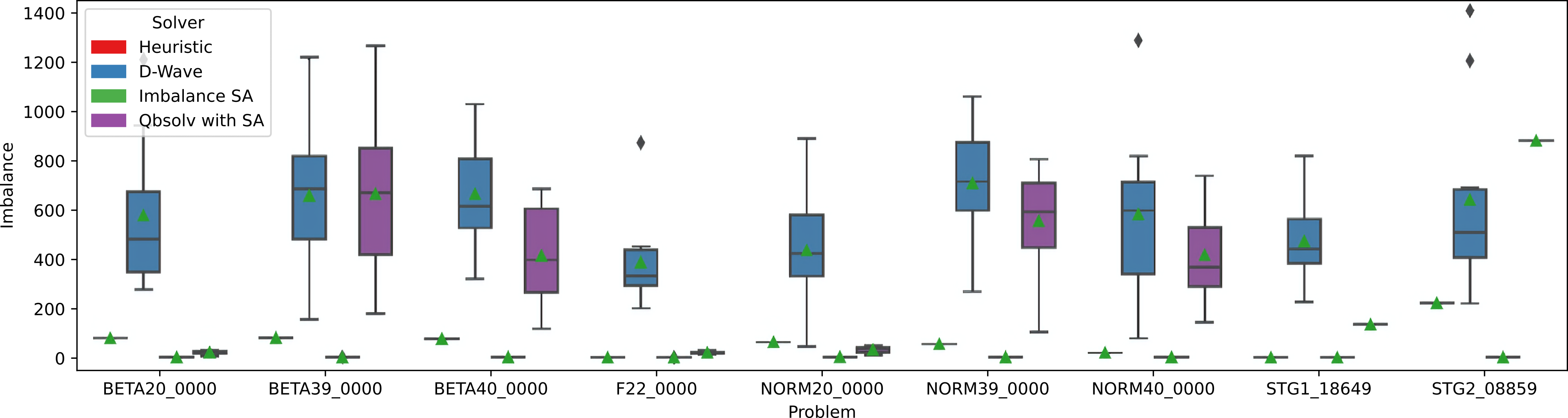}     
    \caption{Total imbalance of the solutions from different solvers: The resulting imbalances from the placement of all blades on the disk are shown. This comparison includes the heuristic, D-Wave QPUs, custom simulated annealing (here denoted as Imbalance SA) and D-Wave's Qbsolv with simulated annealing (from left to right).}
    \label{fig:results_no_init_imbalance}
\end{figure*}

\subsection{QUBO mapping}\label{sec:qubo1}

The TBP has already been transformed into QUBO in~\cite{MASON1997153}, however, this work did not contemplate the bare disk imbalance. 
We will present a formulation including the bare disk imbalance in the following.

Likewise to other permutation-based optimization problems, like the Travelling Salesperson Problem (TSP) with known QUBO formulations~\cite{10.3389/fphy.2014.00005}, we choose a one-hot-encoding basis using $N^2$ binary variables. Instead of considering cities and positions in the cycle as in the TSP, we consider blades with mass $m_i$ attached to slots at angle $\phi_j$. The binary variables $x \in \{0, 1\}^{N \times N}$ encode whether blade $i$ is actually attached to slot $j$, i.e., $x_{ij} = 1$. Therefore, the optimization objective from Eq.~\ref{eq:initial_opt} turns into
\begin{equation}\label{eq:hamiltonian_opt}
\begin{split}
    H(x) &= 2 m_0 \sum_{i=1}^N m_i \sum_{j=1}^N x_{ij} \cos(\phi_0 - \phi_{j})  \\ &+ \sum_{i,k = 1}^N m_i m_j \sum_{j,\ell = 1}^N x_{ij}x_{k\ell}\cos(\phi_{j} - \phi_{\ell}),
\end{split}
\end{equation}
and the solution to $\argmin_x H(x)$ encodes the perfect best mapping of blades. However, so far, the one-hot constraints of the encoding are not considered.

Of course, the above expression can also be written in classical QUBO formalism by $H(x) = x^\top Q x$, with $Q \in \mathbb{R}^{N^2 \times N^2}$. To do so, one has to transform the binary variables into an $N^2$ dimensional vector and introduce the matrix $q = [m_1 \mathbf{z}_1\; m_2 \mathbf{z}_1\cdots m_N \mathbf{z}_N] \in \mathbb{R}^{2 \times N^2}$. Then the QUBO matrix of the Hamiltonian~(as defined in equation~\ref{eq:hamiltonian_opt}) looks like
\begin{align}
    Q = q^\top q + 2 \diag (\mathbf{y}^\top q).
\end{align}

\subsection{Constraints}\label{sec:constraints}

The one-hot constraints can be enforced by introducing Hamiltonians that penalize violations of the constraint. In order for $x$ to represent a permutation matrix, each row and column can only contain a single 1, i.e., one blade can only be positioned at one slot and one slot can only hold one blade. 
Therefore, the Hamiltonian $H_C(x)$ is employed, as defined in equation~\ref{eq:hard_constaints}:
\begin{equation}
\label{eq:hard_constaints}
\begin{split} 
    H_C(x) = \sum_{i = 1}^{N} \lambda_1^{(i)} \left(\sum_{j = 1}^N x_{ij} - 1\right)^2 \\
    + \lambda_2 \sum_{j = 1}^{N} \left(\sum_{i = 1}^N x_{ij} - 1\right)^2,
\end{split} 
\end{equation}
Using $H_C(x)$, every bit-combination that deviates from the one-hot constraint receives a penalty. The penalty factors $\lambda_1^{(i)}\in\mathbb{R}$ and $\lambda_2\in\mathbb{R}$ have to be chosen such that the energy gain in $H$ by violating a constraint is absorbed by the penalty: 
In the case of $\lambda_1^{(i)}$, the penalty term has to be at least $\lambda_1^{(i)} > 2m_0 m_i + m^2_i$ in order to ensure that no blade will be placed more than once.
Here $\lambda_1^{(i)}$ needs to exceed the maximum gain that can be acquired by inserting a blade twice, which arises if the second blade is assigned to the opposing slot to the first blade: $-2m_0m_i - m_i^2$.
In order to guarantee that each slot only contains a single blade, $\lambda_2$ is required to be at least $\lambda_2 > \max_i(2m_0 m_i + m_i^2)$.
This will ensure that even if the heaviest blade is positioned twice, which would yield a maximum gain of $-\max_i 2m_0 m_i + m_i^2$, the hard constraint remains valid.
In our experiments, we set the penalty factors to ten times the minimum requirement based on empirical preliminary tests that have indicated better performance with heavier weighted penalty terms.

Combining all discussed parts, the complete optimization problem can be written as
\begin{align}\label{eq:final_opt}
    \argmin_{x \in \{0, 1\}^{N\times N}} H(x) + H_C(x).
\end{align}

\section{Custom Simulated Annealing Solver}
\label{sec:custom_sa}

Due to the one-hot encoding utilized in our QUBO, it is possible that invalid solutions may occur which do not conform to the physically realizable states. In order to prevent this from happening, we defined the hard constraints in Eq.~\ref{eq:hard_constaints} to ensure compliance. As optimization algorithms mostly find suboptimal results in practice, these constraints may still not satisfied, even though valid solutions yield provably better results. Therefore, we propose an approach to avoid this issue, which we base on the established meta-heuristic simulated annealing. This approach can be tailored towards the optimization of solutions while restricting the solution space to valid solutions.
This coincides with a significant reduction of the search space when compared to the QUBO approach using one-hot encoding.

As part of our work, we have implemented a custom simulated annealing solver, which starts with a random initial solution that satisfies the hard constraints. Based on this initial state, the corresponding energy is calculated by utilizing the function $H(x) = x^\top Q x$, providing a reference value for our optimization. In the next step, the solution is modified by swapping the corresponding one-hot representations of two random blades, ensuring that the hard constraints are always satisfied:
\begin{align*}
    &x = (0\ 0\ 1\ \quad 0\ 1\ 0\ \quad 1\ 0\ 0\ ) \\
    \rightarrow\
    &x = (1\ 0\ 0\ \quad 0\ 1\ 0\ \quad 0\ 0\ 1\ )
\end{align*}
Thereby, the mere exchanging of blades enables us to remain within the valid solution landscape.
The energy is then calculated again for the new candidate and the solution with the minimal energy is preserved. Whether the new solution candidate is considered for future permutations is determined by the applied annealing schedule. For our purposes, we make use of the Metropolis-Hastings algorithm for this schedule~\cite{metropolis1953, hastings1970}.

We repeat this process for a predefined number of iterations $iter \in [1000, 10000]$, after which the best solution will be returned.

\section{Experiments}
\label{sec:experiments}

\begin{figure*}
    \centering
    \includegraphics[width=0.9\textwidth]{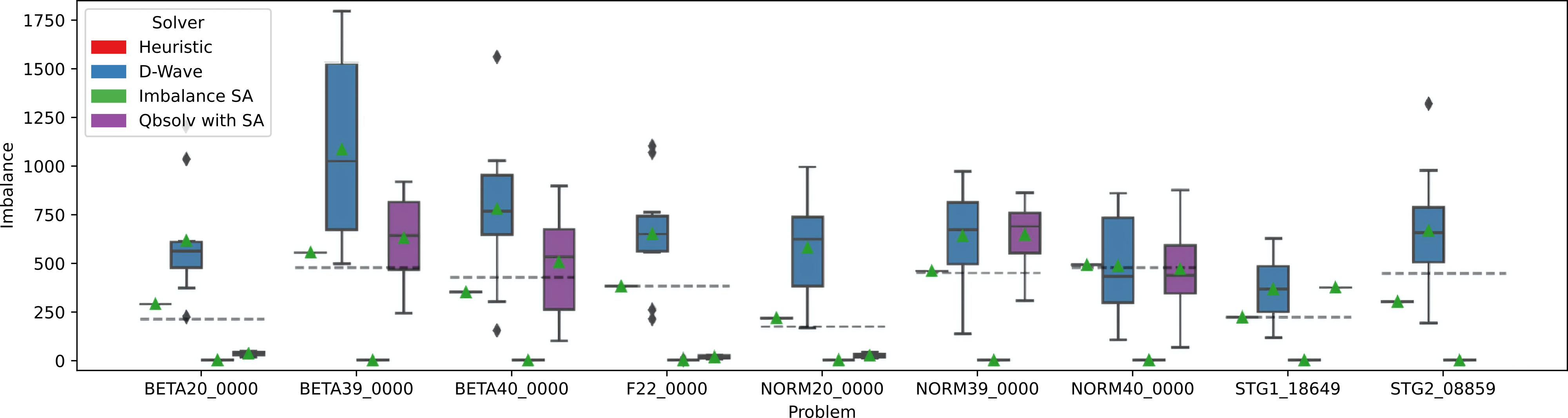}     
    \caption{Total imbalance of the solutions from different solvers (with bare disk imbalances): The resulting imbalances from the placement of all blades on the disk are shown. Additionally, the bare disk imbalance is visualized as a dashed horizontal line. This comparison includes the heuristic, D-Wave's QPU, custom simulated annealing (here denoted as Imbalance SA) and D-Wave's Qbsolv with simulated annealing (from left to right).}
    \label{fig:results_init_imbalance}
\end{figure*}

\label{heuristic_rule}
We used different synthetic and real world datasets, which are directly extracted from an industry use case specifically curated for our experiments.
These were initially solved using an industrially employed classical heuristic.
The chosen heuristic algorithm sorts all of the blades based on their masses and places the two heaviest blades on opposite sides, alternating with the two lightest ones. This procedure is then repeated until all of the blades have been assigned a slot (for a detailed description, see ~\cite{MASON1997153}). 
Note that for our purposes we executed all solvers ten times on each instance to balance statistical relevance with needed computational resources.

\subsection{Dataset}
To allow for meaningful evaluation, we use real world datasets combined with specifically designed synthetic datasets, that aim to capture the fundamentally underlying structure observed real world datasets.
Our synthetic dataset consists of the following instances:
\verb|BETA20_0000|, \verb|BETA39_0000|, \verb|BETA40_0000|, \verb|NORM20_0000|, \verb|NORM39_0000|, and \verb|NORM40_0000|, where each is composed of $20$, $39$, and $40$ blades, respectively.
Here, the \verb|BETA| instances reflect a beta distribution while the \verb|NORM| ones represent a normal distribution of the blade weights. 
In contrast, the \verb|F22_0000|, \verb|STG_18649| and \verb|STG2_08859| instances have been taken from real world datasets consisting of $22$, $84$ and $86$ blades, which exhibit similarly distributed data compared to the synthetic instances.
All synthetic and real datasets have been scaled to show a fixedmean of $10^4$ and a standard deviation of $100$ and are unitless to preserve anonymity.
The underlying distribution of the blade weights is an important factor when solving the respective problem since it may have a significant influence on the strategies employed to obtain a solution.

\subsection{Solvers}

In addition to the heuristic algorithm and the custom simulated annealing solver, we have also applied multiple classical QUBO solvers, such as simulated annealing~\cite{doi:10.1126/science.220.4598.671}, tabu search~\cite{GLOVER1986533}, SAGA~\cite{10.1145/3520304.3534034} and Qbsolv with simulated annealing~\cite{dwavesys-qbsolv} and compared them with each other. 
In the process, we focused on the number of valid solutions of the solvers as the problem size increased.
We found that nearly all of the solvers, beginning at a blade count of $40$, barely returned any valid solutions.
Since Qbsolv showed a comparatively high number of valid solutions and reasonably decent performance for larger problem instances, we will only focus on this solver in the following.
Lastly, we have also used D-Wave's Advantage\_system4.1 QPU in order to solve the balancing QUBOs.
As current hardware does not yet have sufficient resources to handle larger problem instances of the TBP, we have split these problem instances into sub-problems beforehand to be able to solve these on the quantum hardware.
Currently, modern hardware can handle up to $5000$ physical, which corresponds to $119$ fully-connected logical qubits~\cite{dwavesys-tech-report}.
While this enables us to embed a TBP with $13$ blades in princple, the obtained solution quality is rather poor.
One reason for this lies in the hardware noise, which significantly impacts the performance and accuracy ~\cite{Zaborniak_2021}.
As the number of qubits and utilized couplings increases, errors accumulate~\cite{dwavesys-tech-report}.
This noise can result in inaccuracies in the solutions obtained from quantum algorithms, which in turn can degrade the overall performance of the system.
Therefore, based on preliminary tests, the maximum problem size that can be reliably solved on the QPU  is $5$ blades.
To compensate for this constraint, we have divided the problem in two separate groups (assuming the blades are arranged according to the heuristic solution within a list, where each blade is assigned a unique index): one group containing slots and blades with even index numbers, and another group containing slots and blades with odd index numbers.
This partitioning procedure was repeated until each of the sub-problems contained at most $M = 5$ rotor blades.
After solving the sub-problems, the imbalances from the partial solutions were treated as a new TBP and solved, resulting in the complete solution.

\subsection{Results}

The results of the different solvers for the TBP without and with bare disk imbalances are shown in the figures Fig.~\ref{fig:results_no_init_imbalance} and Fig.~\ref{fig:results_init_imbalance}. 
For the case without the additional bare disk imbalance (Fig.~\ref{fig:results_no_init_imbalance}), the heuristic algorithm shows decent performance. However, it is surpassed by the custom simulated annealing solver introduced above in all of the instances.
Likewise, Qbsolv performs on par with the heuristics.
In general, we observe comparably worse solutions with the D-Wave QPU in regard to the resulting imbalances, although it can still outperform the baseline "Heuristic" in some instances, such as \verb|NORM20_0000| or \verb|STG2_08859|, especially when including bare disc imbalances (see Fig.~\ref{fig:results_init_imbalance}).

Note that, as the "Heuristic" solution does not consider the bare disk imbalance, it generally performs worse on these instances.
When examining the solutions obtained by Qbsolv, we observe that in nearly all of the cases, it achieves significantly better solutions than the heuristic.
Meanwhile, for very large problem instances such as \verb|STG1_18649| with $84$ and \verb|STG2_08859| with $86$ blades, Qbsolv performs comparably bad, even finding no valid solution at all for \verb|STG2_08859|.
One possible explanation for this might be that the number invalid solutions grows quadratically with the number of nodes, while the number of valid solutions merely grows linearly.
Not experiencing this shortcoming by definition, the custom simulated annealing presented in Sec.~\ref{sec:custom_sa} yields excellent results for all instances, achieving the industrial requirement for our regarded use case for a maximum total imbalance value of $3$, corresponding to the overall dataset unit scaling, for all cases.
More precisely, the resulting total imbalances from the placement of the blades for the custom simulated annealing solver are merely $1.29 \pm 0.82$.

Finally, examining the D-Wave QPU results, its inferior performance for problem instances without bare disc imbalances is clearly evident.
However, when including bare disc imbalances, as needed in real world application, it performed better then the previously practically employed heuristic.
Nevertheless, when comparing the D-Wave solutions to those produced by Qbsolv and our quantum-inspired approach, it is evident that the latter methods yield significantly improved results in terms of the resulting imbalance.
Consequently no quantum advantage can be claimed when using our setup, i.e., a quantum annealer solving small scale, decomposed QUBO problems.


\section{Conclusion}
\label{sec:conclusion}

This article demonstrates how Quantum Annealing can be used to solve the Turbine Balancing Problem (TBP).
While current quantum hardware already succeeds in surpassing the previously employed heuristic baseline when including the bare disk imbalance, it is outperformed by Qbsolv and the proposed quantum-inspired approach in almost every problem instance.
Correspondingly it must be assessed, that the necessity of formulating the TBP as QUBO in order to be solved with the help of a quantum annealing approach does cause multiple challenges, most importantly, substantial qubit count requirements.
Consequently, limited qubit capacity of current quantum hardware demands employing additional problem decomposition techniques in order to properly handle medium and large scale real world TBP instances.
Furthermore, strong evidence was given, that the ratio between valid and invalid solutions, caused by the representation of the problem as a QUBO, introduces additional obstacles.
These issues led to the proposal of a novel, quantum-inspired approach, which successfully solves the TBP to the extent that it meets the industry requirements for all instances.
This in turn raises prospects of finding an effective way to solve this problem utilizing quantum computing when reducing the search-space to only-valid solutions, by employing binary encoding instead of one-hot encoding. As this leads to a PUBO rather than a QUBO problem formulation, quantum optimization algorithms beyond quantum annealing (such as the QAOA) should be investigated for solving the TBP in future work.



\bibliographystyle{ACM-Reference-Format}
\bibliography{references}




\end{document}